# Online Suicide Games: A Form of Digital Self-harm or A Myth?

Maria BADA[a,1] and Richard CLAYTON [a]
[a]*Computer Laboratory, University of Cambridge, Cambridge, UK Affiliation*

**Abstract.** Online 'suicide games' are claimed to involve a series of challenges, ending in suicide. A whole succession of these such as the Blue Whale Challenge, Momo, the Fire Fairy and Doki Doki have appeared in recent years. The 'challenge culture' is a deeply rooted online phenomenon, whether the challenge is dangerous or not, while social media particularly motivates youngsters to take part because of their desire for attention. Although there is no evidence that the suicide games are 'real', authorities around the world have reacted by releasing warnings and creating information campaigns to warn youngsters and parents. We interviewed teachers, child protection experts and NGOs, conducted a systematic review of historical news reports from 2015-2019 and searched police and other authority websites to identify relevant warning releases. We then synthesized the existing knowledge on the suicide games phenomenon. A key finding of our work is that media, social media and warning releases by authorities are mainly just serving to spread the challenge culture and exaggerate fears regarding online risk.

**Keywords.** Suicide, self-harm, online games, prevention

## 1. Introduction

Much attention is currently given to the threat apparently posed by online suicide games, such as the Blue Whale Challenge, Momo, the Fire Fairy, Doki Doki, and others. They are said to involve posting about progress through a series of challenges, which include self-harm behaviour, ending in suicide. Authorities regularly draw attention to these games by issuing warnings about them.

The 'challenge culture' is a deeply rooted online phenomenon. Besides suicide games there are many other less obviously dangerous challenges such as the salt and ice challenge, the cinnamon challenge and more recently skin embroidery. Regardless of whether a challenge is dangerous or not, youngsters are especially motivated to take part, presumably because of a desire for attention. They are constantly faced with a barrage of cultural norms and values, what their peers think and do, media coverage and social policies, each of which inform their thinking, values and beliefs [1].

Mainstream media links the Blue Whale Challenge to at least 130 teen deaths in Russia and attempts at suicide in Spain and Ukraine [2]. However, despite these claims there is no evidence for large numbers of deaths linked to suicide games. Snopes investigated Blue Whale in 2017 and deemed the story 'unproven' [3]. In 2019 the BBC posted a detailed history of Blue Whale showing there was no record of such a game prior to a single Russian media article of dubious accuracy [4].

---

[1] Author: Maria Bada, Computer Laboratory, University of Cambridge, Cambridge, UK; E-mail: maria.bada@cl.cam.ac.uk

The UK Safer Internet Centre (https:\\saferinternet.co.uk) calls the claims around Momo 'fake news', while YouTube has (despite this being central to media accounts) found no evidence of videos showing or promoting Momo on its platform. Researchers who studied five Italian case histories allegedly linked to Blue Whale found that the patients, although self-harming, were not participating in a suicide game [5]. Nevertheless, authorities around the world continue to react to inaccurate media reports by releasing warnings and creating information campaigns to warn youngsters and parents about the risks related to these online suicide games.

This paper examines the online phenomenon of the 'challenge culture' and its association with self-harm, focussing on: 1) identifying the social media spread of the challenge culture; 2) understanding the behaviour of young people taking interest in the Blue Whale challenge and other games; and 3) exploring the practices of authorities in providing warnings and awareness programmes.

## 2. Literature Review

*2.1. Self-harm*

Previous research shows that the Internet is spreading self-harm behaviour among vulnerable teenagers who are characterised by epidemiological, psychological, psychiatric, social, and cultural risk factors [6]. However, there is little research on the reasons or the motivation of young people for revealing self-harm online.

Boyd [7] speculated on three possibilities: self-harmers might be uttering a cry for help, they might want to appear 'cool', or they may be trying to trigger compliments. If digital self-harmers are 'crying for help', this indicates that mental health problems or disorders might be present. Self-harm is the strongest predictor of suicide among young people and adolescents who self-harm report that it is difficult to talk about their suicidal feelings and that they do not really feel 'listened to' when they do communicate on this topic [5]. Research has shown that self-harm and depression are linked to an increased risk of suicide and so, online manifestation of self-harm behaviours may precede suicide attempts [8]. Self-harm may be a demonstration of felt pain and distress while those who participate in offline self-harm are more likely to be involved in digital self-harm [9].

Studies suggest that the Internet may exacerbate existing risks, illustrated by examples of individuals who commit to killing themselves in online suicide for a and then feel they cannot back down [10]. Phippen [11] stresses that both content and behaviour have a contribution to causing upset online for young people. Young people use online technology for communication, they do not merely consume content with these platforms.

*2.2. Technopanics*

The term 'technopanic', often used by experts in child online safety, highlights the exaggeration of sometimes real, other times non-existent, risks in the digital world. Technopanic is not new, it has accompanied all new technology - from electricity, to radio and television. But digital technopanic has a special characteristic which amplifies

it - the submerging of the new generation in new technologies and the unpreparedness of parents for 'digital parenting' [12]. Overcoming fear may be perceived as adventurous and psychologically rewarding. It is natural that daring against an unexposed fear gives a feeling of pleasure and satiation, especially upon success. Adventure games exploit this fear psychology [13]. So exaggerating youth risk-taking and spreading both the hype and fear only increases the risk to young people who are relying on perception rather than facts. It directly impacts their behaviour, in this case in a negative direction [14].

*2.3. Media Literacy*

Understanding what media content really means, what its source is and why a certain message has been constructed, is crucial for quality understanding and recognition of media mediated messages and their meaning. Adequate answers to all these questions can only be acquired by media literacy [15]. However, in most countries media education is still a secondary activity that teachers or media educators deal with without training or proper material. It has been added to the curriculum in many countries but much more is needed in practice.

A key concern when discussing suicide games is the risk of 'suicide contagion' [16], which could turn stories into a tragic self-fulfilling prophecy for a small number of vulnerable youths. The American Foundation for Suicide Prevention has published guidelines for journalists that include avoiding, "*big or sensationalistic headlines, or prominent placement of suicide reports, and not describing recent suicides as an 'epidemic', 'skyrocketing', or other strong terms*" [17].

*2.4. Policy Implications*

As a response to incidents of youth suicide and the linkage of these incidents to online suicide games, policy decisions have been made in the UK and other countries. In the UK social media firms are now being forced to disclose data that could reveal their role in fuelling self-harm and suicides [18]. The UK Government have now published a White Paper [19] on online harms which sets out the government's plans for a package of measures to keep UK users safe online.

There have been also national level policy decisions such as taking down websites. In India, this has led to a court decree for shutting down websites to stop the spread of the Blue Whale challenge [20]. The Indian government has also asked companies like Google, Facebook, WhatsApp, Instagram, Microsoft and Yahoo to remove all links related to Blue Whale Challenge [20]. Additionally, in 2017 Russia passed a law [21] introducing criminal responsibility for creating pro-suicide groups on social media and for inducing minors to commit suicide.

## 3. Research Methods and Results

We took several approaches to this research, so as to triangulate data from different sources. Triangulation refers to using multiple data sources in qualitative research to develop a comprehensive understanding of phenomena [22]. Triangulation has also been viewed as a qualitative research strategy to test validity through the convergence of information from different sources. Below we present the results from a) interviews; b) news reports and media; and c) warning releases.

*3.1. Interviews with Teachers, Child Protection Experts & NGOs*

We conducted five interviews with teachers, child protection experts and NGOs using open-ended questions. The questions used referred to: a) the challenge culture; b) youth and self-harm; c) psychological disorders associated with self-harm; d) the role of parents and schools; e) the links to authorities; f) awareness programmes or training provided to teachers and parents and g) ICT-risk related policies at schools.

We found a lack of understanding that online suicide games are 'fake news' and a general misperception of how to react to related incidents. A teacher informed us that "*Recently, two teachers from two different classes heard children talk about Momo, they were planning to call over the weekend* (Participant 1)." The school then decided to release a warning email to parents.

Additionally, interviewees informed us that "*according to school policy children under 13 years old, are not allowed to use social media, and if they do it needs to be reported* (Participant 1)." According to interviewees, "*children think that it is cool to discuss about these games, they would play a game because their friends do and then they would talk about it at school. It is cool, it is trendy* (Participant 2)."

Regarding the 'challenge culture', an interviewee mentioned that "*children do enjoy the YouTube videos and anything that has to do with challenge and then there is also peer pressure* (Participant 3)." It was also mentioned that "*Facebook and Google have tried to control the circulation of videos with the title challenge, Momo etc. but people would find a way around it and still circulate videos* (Participant 3)."

An interviewee from an NGO noted that "*media started to repeat the story about the Blue Whale Challenge back in 2015. At that point the helpline was receiving 35 calls every day, people were worried, and the story spread quickly. Schools also shared warnings with parents. Then parents shared with Facebook. However, these were no real case evidence* (Participant 4)." An expert from an NGO further noted, "*There is no suicide game, no risk. The real risk is lack of communication and the generation divide. If we talk to a specific challenge or game, then it promotes the game. We need to talk in general* (Participant 5)."

*3.2. Media reports review*

We searched for news reports in the period 2015{2019 using the keywords: 'Online AND Suicide AND Game', 'Self-harm', 'Blue Whale Challenge', 'Momo Challenge', 'Blue Whale Suicide Game', 'Doki Doki'. We then used Buzzsumo to count articles, Twitter shares and Facebook mentions in the last two years.

As Table 1 shows, the original news stories have been amplified many times through social media with news about Momo spreading further than the Blue Whale challenge. In February 2019, Momo was once again picked up by UK media and social media, leading to a rapid increase of search interest from the public [23].

These findings indicate that seeing articles or news reports of online suicide games is becoming more common. The main issue here is whether media reporting on the online suicide games is becoming the norm. As mentioned above, there is a general misperception and a general lack of media literacy around this topic.

**Table 1.** References (blogs, tweets etc.) to relevant media articles 2017-2019

| Media source 'Blue Whale' | References | Media source 'Momo' | References |
|---|---|---|---|
| The Sun | 8.479 | BBC | 718.500 |
| BBC | 7.700 | The Atlantic | 157.300 |
| Daily Mail | 3.400 | Forbes | 107.900 |
| | | The Guardian | 84.700 |
| | | New York Times | 31.000 |
| | | Wired | 23.700 |
| | | National Online Safety | 20.000 |
| | | The Telegraph | 1.500 |

*3.3. Police and School Warning Releases*

We searched UK Police and related websites for warnings about online suicide games. We found some use of the web to post warnings to alert parents, but that warnings often appeared on social media such as Twitter or Facebook.

A Devon and Cornwall Police PCSO posted on Twitter "*Who ever created this horrible game is sick! Parents: Please be aware of this 'game' talk to your children about it if concerned*" [2] (Quote 1) and Cambridgeshire Police warned parents about "*a sinister social media suicide game*" (Quote 2) but both said there were no reports of anyone in their counties taking part in the game [24].

Schools have also been asked to inform parents. Schools in Essex, Cambridgeshire, Hertfordshire, Cornwall and elsewhere have sent letters to parents explaining the risks involved in the challenges. A central justification appears to be that "*the school has a duty of care for young people*" [25, 26].

**4. Discussion**

News reports and articles about online suicide games are extremely common, but there is a lack of media literacy around this topic which means that these reports are often taken at face value, whereas the overwhelming evidence is that they are 'fake news'. This is compounded by a lack of knowledge about how best to report suicide incidents so as to avoid any possibility of 'contagion'.

We found that police officers in Britain, France and Belgium have been posting warnings about suicide games using media such as Twitter and Facebook. However, Bulgaria's Centre for Safe Internet reported in 2016 that stories about suicide games were essentially online rumours and no basis in fact could be found. The Centre made the link to a more general inability of children and their parents to distinguish fake news from real content on the Internet [27].

Better media literacy is needed, but training and awareness-raising programmes need to be frequent and tailored to the needs and interests of each group and communicated in a way and language they can easily relate with [28]. Different approaches are available to support young people such as a games, videos or exercise-based educational material [29]. Our research leads us to the following policy recommendations:

a) **Awareness and education** to ensure that young people can handle risks online and offline [28]. Training of teachers is needed as well as study materials. Online media education must take its place in the curriculum [30], but considerable attention also needs to be devoted to helping and empowering parents, so that they can guide their children to personally determine if an activity is risky or not;
b) **Guidelines for media and social media reporting of suicides** are needed at national and international levels [17] to ensure that media and social media report suicides responsibly. These guidelines must evolve in line with improved strategies for suicide prevention and the development of community response plans;
c) **Improve social media and media understanding of suicide**. Media must avoid re-running details of each death in every report, avoid speculation about the 'trigger' for a suicide and should use the story to inform readers about the causes of suicide, its warning signs, trends in rates and recent treatment advances [31];
d) **Collaborative efforts to prevent 'fake news' about suicides**. Google, Facebook and Twitter have already floated a range of ideas to combat the spread of fake news more generally, including compiling lists of fake news sites, flagging certain stories as having been disputed as fake, using plug-ins and apps to detect fake news, and even taking down known fake news providers [32];
e) **Quality control of warning releases by authorities**. Authorities must follow general guidelines for news about suicides, warning but not stimulating extra interest. However, they must go further to ensure that they are not just another source of fake news. They must collaborate with other authorities and with NGOs to ensure that their warnings relate to real events - and then, when appropriate, work with cybercrime units to deal with criminal behaviour.

We do not dispute that suicide amongst young people is of justifiable concern, but we can find no evidence of suicide games leading to mass suicide in the way that the hype suggests. Media and social media are copying stories about suicide games and feeding on each other with little evidence that their tales are true. The exaggerated fears around online risk and official warnings merely promote the challenge culture and drives youngsters to seek out these challenges.